 \documentclass[preprint2]{aastex}

\usepackage{array}
\usepackage[figuresright]{rotating}
\usepackage{booktabs}
\usepackage{txfonts}

\shorttitle{The Nearest High-Velocity Stars Revealed by LAMOST Data Release 1}
\shortauthors{J. Zhong et al.}
\begin{document}


\title{The Nearest High-Velocity Stars Revealed by LAMOST Data Release 1}



\author{Jing Zhong\altaffilmark{1}, Li Chen\altaffilmark{1}, Chao
  Liu\altaffilmark{2}, Richard de Grijs\altaffilmark{3}, Jinliang
  Hou\altaffilmark{1}, Shiyin Shen\altaffilmark{1}, Zhengyi
  Shao\altaffilmark{1}, Jing Li\altaffilmark{1}, Ali
  Luo\altaffilmark{2}, Jianrong Shi\altaffilmark{2}, Haotong
  Zhang\altaffilmark{2}, Ming Yang\altaffilmark{2}, Licai
  Deng\altaffilmark{2}, Ge Jin\altaffilmark{5}, Yong
  Zhang\altaffilmark{6}, Yonghui Hou\altaffilmark{6}, and Zhenchao
  Zhang\altaffilmark{6}}

\altaffiltext{1}{Key Laboratory for Research in Galaxies and
  Cosmology, Shanghai Astronomical Observatory, Chinese Academy of
  Sciences, 80 Nandan Road, Shanghai 200030, China; jzhong@shao.ac.cn}

\altaffiltext{2}{Key Laboratory of Optical Astronomy, National
  Astronomical Observatories, Chinese Academy of Sciences, Datun Road
  20A, Beijing 100012, China}

\altaffiltext{3}{Kavli Institute for Astronomy \& Astrophysics and
  Department of Astronomy, Peking University, Yi He Yuan Lu 5, Hai
  Dian District, Beijing 100871, China}

\altaffiltext{5}{University of Science and Technology of China, Hefei
  230026, China}

\altaffiltext{6}{Nanjing Institute of Astronomical Optics and
  Technology, National Astronomical Observatories, Chinese Academy of
  Sciences, Nanjing 210042, China}

\begin{abstract}
We report the discovery of 28 candidate high-velocity stars (HVSs) at
heliocentric distances of less than 3 kpc, based on the Large Sky Area
Multi-Object Fiber Spectroscopic Telescope (LAMOST) Data Release
1. Our sample of HVS candidates covers a much broader color range than
the equivalent ranges discussed in previous studies and comprises the
first and largest sample of HVSs in the immediate solar neighborhood,
at heliocentric distances less than 1--3 kpc. The observed as well as
the derived parameters for all candidates are sufficiently accurate to
allow us to ascertain their nature as genuine HVSs, of which a subset
of 12 objects represents the most promising candidates. Our results
also highlight the great potential of discovering statistically large
numbers of HVSs of different spectral types in LAMOST survey
data. This will ultimately enable us to achieve a better understanding
of the nature of Galactic HVSs and their ejection mechanisms, and to
constrain the structure of the Galaxy.
\end{abstract}


%
\section{Introduction}           
\label{sect:intro}

Hypervelocity stars (HVSs) move sufficiently fast so that they can
escape from the Galaxy's gravitational pull. They may have attained
their high velocities through three-body interactions, for instance
among binary systems in star clusters \citep[e.g.,][and references
  therein]{leonard91,moy13} or with the supermassive black hole in the
Galactic Center \citep[GC;
  e.g.,][]{hills88,yu03,2005ApJ...622L..33B}. While this latter
formation mechanism is thought to be very promising, other HVS
ejection scenarios are possible, such as those involving close
encounters of single stars with binary black holes \citep{yu03} or the
disruption of stellar binaries in the Galactic disk
\citep{2009A&A...508L..27W,napiwotzki.2012}.

Until recently, most confirmed and potential HVSs had been
  identified as early-type stars located at Galactocentric distances
  greater than 20 kpc. At present, dozens of confirmed HVSs are known
  in the Milky Way (\citealt{2005ApJ...622L..33B, 2009ApJ...690.1639B,
    2012ApJ...751...55B}; see also \citealt
  {2009A&A...507L..37T,tillich10}), most of which are massive B-type
  stars. To distinguish among HVS ejection mechanisms and place firm
  constraints on the origin of their parent population, construction
  of larger HVS candidate samples spanning a much wider range of
  spectral types is imperative. Based on Sloan Digital Sky Survey
  (SDSS) data, \citet{2009ApJ...697.1543K,2010ApJ...723..812K}
  attempted to find metal-rich, old-population HVS stars (mainly F/G
  stellar types) that had been ejected from the GC. However, their
  non-detection of such old-population ejectees only places a limit on
  the rate of ejection, which suggests that the stellar mass function
  in the GC may be top- instead of bottom-heavy, or alternatively that
  the supermassive black hole ejection mechanism often involed for the
  ejection of lower-mass stars is less effective than expected.
  Recently, \citet{2012ApJ...744L..24L} reported the discovery of 13
  F-type HVSs, located at distances ranging from 3 kpc to 10
  kpc. \citet{palla14} identified 20 high-velocity G- and K-dwarf
  candidates, discovered in the Sloan Extension for Galactic
  Understanding and Exploration (SEGUE) sample, most of which are
  located far beyond the solar neighborhood (at distances of 3--6
  kpc). The first HVS discovered in the Large Sky Area Multi-Object
  Fiber Spectroscopic Telescope (LAMOST) survey, a B-type star
  characterized by a galactrocentric radial velocity component of
  approximately 477 km s$^{-1}$ and a heliocentric distance of
  $\sim$13 kpc, was recently reported by \citet{2014ApJ...785L..23Z}.

Here we report the discovery of 28 solar-neighborhood HVS candidates
based on the LAMOST Data Release 1 (DR1). All candidates have
velocities, with respect to the Galactic rest frame, in excess of 300
km s$^{-1}$; 12 objects move faster than 400 km s$^{-1}$. Our HVS
candidate sample covers a much more extended color range than
addressed in previous studies, while most targets are located at
heliocentric distances within 3 kpc (some even within 1 kpc). Our
sample comprises the first large HVS sample in the immediate solar
neighborhood. As a consequence, it is of great importance for
investigating the various HVS ejection mechanisms, which may be more
complex than previously thought \citep[e.g.,][]{palla14}.

This {\it Letter} is organized as follows. In Section~\ref{sect:data},
a brief description of the LAMOST survey data and our candidate
selection is presented. We discuss distance estimates to our HVS
candidates and derive their kinematic properties in
Section~\ref{sect:Select}. The key results are presented in Section
\ref{sect:discussion}.


\section{LAMOST Survey Data and HVS Candidate Selection}
\label{sect:data}

\subsection{LAMOST DR1 Data}

LAMOST---the ``Guoshoujing telescope''---is a Schmidt telescope with
an effective aperture of 4 m. It is equipped with 4000 fibers that can
be deployed across a 5$^\circ$ (diameter) field of view, with a
spectral resolution of $R \simeq 1800$ over the wavelength range from
3800 {\AA} to 9100 {\AA} \citep{cui12,zhao12}.

DR1, based on a year of pilot operations and the first season of
official survey science, contains $1.7 \times 10^6$ spectra suitable
for accurate radial-velocity (RV) measurements. Most of these spectra
represent observations obtained as part of the LAMOST Experiment for
Galactic Understanding and Exploration (LEGUE) survey \citep{deng12}
aimed at unveiling the structure of the Milky Way. A subset of some
1.3 million spectra, mostly of bright ($r < 17$ mag) stars spanning a
broad color range, are characterized by signal-to-noise ratios (S/N)
greater than 5 in the SDSS $g$, $r$, and $i$ filters. The sample
selection criteria
\citep[cf.][]{carlin12,2012RAA....12..805C,yang12,zhang12} favor
inclusion of nearby main-sequence stars, simply because of their
preponderance along any line of sight.

\subsection{HVS Candidate Selection}

\begin{figure*}
   \centering
   \includegraphics[angle=0,scale=.4]{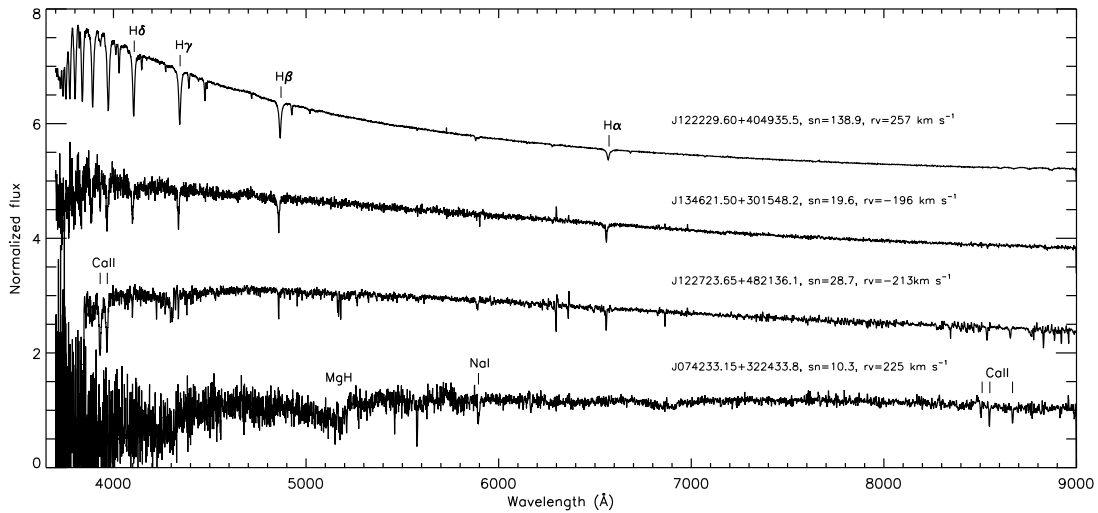}
   \caption{Typical spectral sequence representative of our HVS
     candidates. A number of different absorption lines are used for
     measuring the average wavelength shift and computing the RVs. The
     LAMOST designation, S/N, and RV are shown above the respective
     spectrum. }
   \label{Fig1}
\end{figure*}

By 2013 March 20, regular LAMOST survey operations had observed and
processed around one million stellar spectra. The standard reduction
pipeline \citep{zhao12} converts two-dimensional into one-dimensional
(1D) spectra, applies flat-field corrections, combines the blue and
red spectral ranges, and takes care of wavelength calibration and sky
subtraction. The pipeline also provides stellar RVs based on cross
correlation with the ELODIE library \citep{prug07}. Although the
process of stellar parameter determination, including of the
appropriate stellar surface gravities ($\log g)$, effective
temperatures ($T_{\rm eff}$), and metallicities ([Fe/H]), is still
being refined, stellar RVs based on ``good-quality spectra'' (S/N $>
5$) are highly accurate \citep[cf.][where the quotes refer to verbatim
  text from that reference]{luo12}, with uncertainties of order 13 km
s$^{-1}$.

For a systematic investigation of HVS candidates, we first selected
the 14,650 stars in the LAMOST DR1 with absolute RV values greater
than 200 km s$^{-1}$. Next, to derive three-dimensional (space)
velocities, we identified the 6148 targets for which we had both
proper motions---from the Fourth U.S. Naval Observatory CCD Astrograph
Catalog (UCAC4)---and SDSS DR9 photometry ($r < 18.5$ mag) available,
with sufficient S/N in both the $g$ and $i$ bands (S/N $\ge 6$ and S/N
$\ge 20$, respectively). This resulted in an astrometric catalog
containing 2600 candidates.

To derive photometric distances, we need reliable estimates of the
stellar metallicities. Starting from our 2600 candidate HVSs, we
further excluded stars for which either their LAMOST-based [Fe/H]
values are not available or their SDSS colors fall outside either the
$[ 0.2 < (g-r) < 0.6, 0.8 < (u-g) < 1.3 ]$ mag or $[ 0.1 < (r-i) < 1.6
]$ mag ranges \citep{2008ApJ...684..287I}, where sufficiently accurate
photometric metallicities can be derived. We subsequently selected the
remaining 1800 candidates to calculate the preliminary stellar
distances and space velocities in the Galactic rest frame ($V_{\rm
  gt}$). Upon imposing an additional selection criterion of $V_{\rm
  gt} > 300$ km s$^{-1}$, only approximately 200 candidates were
left. We visually inspected this final subsample and calculated the
equivalent widths of the Mg${b}$ (5180 {\AA}), H${\beta}$ (4863
{\AA}), and TiO lines (6150 {\AA}) for each LAMOST spectrum. We next
verified the stellar RVs in our sample catalog and excluded those
spectra that were characterized by low S/N. The RVs resulting from
such observations were insufficiently reliable because they lacked a
sufficient number ($\ge 3$) of spectral lines for calibration. We also
excluded those spectra that more likely represented the
characteristics of red giant stars according to the classification
criteria derived by \citet{Liu14}. We finally obtained a HVS candidate
sample containing 28 objects: see Table 1.

\section{HVS Candidate Verification}
\label{sect:Select}

\subsection{RV Verification}
\label{sect:rvver}

To verify the LAMOST RV results, we used the
IRAF/\texttt{rv.rvidlines} package. Figure~\ref{Fig1} shows four
typical spectra representative of our sample objects. For early-type
stars (top and second panels), we predominantly used the Balmer
lines---H${\alpha}$ (6563 {\AA}), H${\beta}$, H${\gamma}$ (4341
{\AA}), and H${\delta}$ (4102 {\AA})---to compute the average
wavelength shift and obtain the best-fitting RV; for late-type stars
(third and bottom panels), we used other appropriate lines for
calibration, including the calcium Ca {\sc ii}) H and K lines at,
respectively, 3933 {\AA} and 3968 {\AA} and the near-infrared
(near-IR) (Ca {\sc ii} triplet at 8498 {\AA}, 8542 {\AA}, and 8662
{\AA}. The mean RV errors for all spectra were less than 10 km
s$^{-1}$.

Among our data set of 28 promising HVSs, five also have SDSS RV
measurements. For the latter stars, we found reasonable consistency in
the RV results. The residual standard deviation ($\sigma$) was 16.8 km
s$^{-1}$ when comparing our RV results with the SDSS DR9 values, and
$\sigma =29.7$ km s$^{-1}$ when comparing the LAMOST 1D pipeline
results with the SDSS data. This also implies that our IRAF reduction
is more reliable than the LAMOST 1D pipeline results.  Below
  (e.g., in Section \ref{sect:discussion}), we compare the properties
  of our candidate HVSs in the context of literature-based
  determinations, where available, in more detail.

Note that a well-known, hyper-runaway B-type star \citep[HIP 60350;
  see
  also][]{1998A&A...339..782M,2001A&A...369..530T,2010ApJ...711..138I},
is also included in our initial sample catalog (LAMOST designation:
J122229.60+404935.5; see the top panel of Figure~\ref{Fig1}). This
star's RV is $248 \pm 13$ km s$^{-1}$ based on the LAMOST 1D pipeline
and $257 \pm 8$ km s$^{-1}$ as derived from our IRAF results. This is
in agreement with \citet{2010ApJ...711..138I}, who found a barycentric
RV of $v_{\rm rad}=262 \pm 5$ km s$^{-1}$, based on a high-resolution
optical echelle spectrum.

\begin{table*}
\caption{Basic Parameters of the 28 HVS Candidates.}
\label{tab1}
\tiny
\begin{tabular}{cccrrrrrrrrrrrrrrr}
\\
\hline
 Designation &R.A. & Dec.& $~~~~~~$$\mu_{\alpha} \cos(\delta)$ & $\mu_{\delta}$$~~~~~~$ & $u$$~~~$ &   $g$$~~~$ & $r$$~~~$ & $i$$~~~$  &  $z$$~~~$  & $J$$~~~$ & $H$$~~~$ & $K_{\rm s}$$~~$& $W2$$~~$ & [Fe/H] & $V^{a}_{\rm r}$$~~~~$ & $V^{b}_{\rm r}$$~~~~$ & SpTy\\
  $~$ & deg & deg & $~~~~~$mas yr$^{-1}$  & $~~~~~$mas yr$^{-1}$ & mag$~$ & mag$~$ & mag$~$ & mag$~$ & mag$~$ & mag$~$ & mag$~$ & mag$~$ &  mag$~$ & $~$dex & km s$^{-1}$ & km s$^{-1}$ & $~$\\
\hline
   J085819.90+150352.4 &  134.58293 & 15.064578 &  $-$21.8 $\pm$  2.6&  $-$54.3 $\pm$  3.0  &  15.23 & 14.22 & 14.38 & 14.13 &  13.82 &  12.94 &  12.69 &  12.64 &  12.63 &  $-$1.11 &   $~~$428$~~~~$   &   $~~$415$~~~~$    &  A5\\
   J092707.09+242752.9 &  141.77954 & 24.464709 &  $-$59.8 $\pm$  8.5&  $-$2.7 $\pm$  8.0  &  17.17 & 16.18 & 15.83 & 15.90 &  15.67 &  14.90 &  14.58 &  14.50 &  14.40 &  $-$1.56 &   $~~$306$~~~~$   &   $~~$308$~~~~$     &  F5\\
   J074233.15+322433.8 &  115.63816 & 32.409389 &  $-$156.0 $\pm$  8.0&     $-$353.0 $\pm$  8.0  &  19.96 & 17.47 & 16.12 & 15.57 &  15.26 &  14.10 &  13.62 &  13.43 & 13.09 &---$~~$ &   $~~$233$~~~~$   &   $~~$225$~~~~$  &  K7\\
   J225922.58+262409.9 &  344.84410 & 26.402775 &  $-$49.3 $\pm$  4.2&  72.2 $\pm$  4.5  &  16.19 & 15.26 & 14.96 & 14.84 &  14.81 &  14.04 &  13.70 &  13.64 &  13.62 &---$~~$ &   $-$229$~~~~$   &  $-$220$~~~~$            &  B6\\
   J115445.93+523520.3 &  178.69142 & 52.588984 &   56.3 $\pm$  2.2&  $-$76.4 $\pm$  2.6  &  16.42 & 15.36 & 14.95 & 14.90 &  14.75 &  13.88 &  13.55 &  13.53 &  13.46 &---$~~$ &   $-$228$~~~~$   &  $-$219$~~~~$           &  F2\\
   J082932.48+282613.3 &  127.38536 & 28.437039 &    $-$196.0 $\pm$  8.0&     $-$395.0 $\pm$  8.0  &  18.99 & 16.51 & 15.18 & 14.64 &  14.34 &  13.20 &  12.64 &  12.44 &  12.21 &---$~~$ &  $~~$279$~~~~$  &  $~~$282$~~~~$  &  K7 \\
   J231232.05+265039.5 &  348.13358 & 26.844326 &   63.6 $\pm$  2.9&  14.6 $\pm$  3.0  &  16.12 & 15.08 & 14.63 & 14.78 &  14.37 &  13.48 &  13.08 &  13.00 &  13.01 &  $-$0.70 &   $-$235$~~~~$   &  $-$196$~~~~$            &  F5 \\
   J135931.65+413121.6 &  209.88189 & 41.522689 &  $-$70.3 $\pm$  2.8&  $-$78.4 $\pm$  3.7  &  16.47 & 15.47 & 15.11 & 14.96 &  14.93 &  14.11 &  13.75 &  13.66 &  13.70 &---$~~$ &   $~~$261$~~~~$   &   $~~$244$~~~~$      &  F5 \\
   J134621.50+301548.2 &  206.58962 & 30.263401 & $-$15.2 $\pm$  3.9&  $-$58.4 $\pm$  4.2  &  16.56 & 15.66 & 15.45 & 15.37 &  15.37 &  14.59 &  14.37 &  14.37 &  14.21 &---$~~$ &   $-$244$~~~~$   &  $-$196$~~~~$          &  A0 \\
   J110208.65+575200.2 &  165.53608 & 57.866742 &   70.6 $\pm$  1.9&  $-$84.0 $\pm$  1.6  &  14.67 & 13.42 & 14.61 & 12.95 &  13.24 &  11.96 &  11.65 &  11.63 &  11.61 &  $-$1.16 &   $-$219$~~~~$   &  $-$227$~~~~$         &  A0 \\
   J224145.15+292428.4 &  340.43813 & 29.407892 & $-$15.3 $\pm$  7.2&      $-$60.4 $\pm$  10.0 &  16.03 & 15.09 & 14.88 & 14.78 &  14.76 &  13.99 &  13.72 &  13.72 &  13.73 &---$~~$ &   $-$220$~~~~$   &  $-$210$~~~~$      &  A0 \\
   J124020.02+453259.6 &  190.08345 & 45.549896 &  26.2 $\pm$  2.6&  $-$52.8 $\pm$  2.6  &  16.32 & 15.42 & 15.15 & 15.05 &  15.06 &  14.34 &  14.01 &  14.02 &  13.95 &  $-$1.30 &   $-$219$~~~~$   &  $-$195$~~~~$          &  A0 \\
   J123134.97+424736.1 &  187.89571 & 42.793379 &  17.2 $\pm$  2.9&  $-$49.2 $\pm$  2.6  &  16.83 & 15.89 & 15.57 & 15.46 &  15.44 &  14.59 &  14.31 &  14.18 &  14.36 &  $-$1.35 &   $-$254$~~~~$   &  $-$266$~~~~$          &  F0 \\
   J094122.37-000822.2 &  145.34323 &$-$0.139505 &$-$46.5 $\pm$  4.4&  $-$67.6 $\pm$  4.6  &  16.54 & 15.54 & 15.23 & 15.07 &  15.05 &  14.20 &  14.01 &  13.91 &  13.85 &  $-$1.62 &   $~~$430$~~~~$   &   $~~$426$~~~~$     &  F0 \\
   J073050.71+293212.6 &  112.71130 & 29.536856 &   21.9 $\pm$  3.2&  $-$77.7 $\pm$  2.9  &  15.85 & 14.86 & 14.52 & 14.40 &  14.37 &  13.56 &  13.21 &  13.22 &  13.16 &  $-$1.20 &   $-$309$~~~~$   &  $-$291$~~~~$         &  F5 \\
   J123641.48+443715.9 &  189.17285 & 44.621092 &  $-$23.7 $\pm$  3.2&  $-$55.6 $\pm$  3.0  &  17.22 & 16.35 & 16.07 & 15.95 &  15.95 &  15.12 &  14.90 &  14.95 &  14.79 &  $-$1.79 &   $-$257$~~~~$   &  $-$259$~~~~$       &  F0 \\
   J085623.74+211322.3 &  134.09893 & 21.222872 &  $-$60.0 $\pm$  3.4&  $-$47.3 $\pm$  2.6  &  15.09 & 14.15 & 13.81 & 13.88 &  13.77 &  12.83 &  12.50 &  12.51 &  12.48 &  $-$1.46 &   $~~$223$~~~~$   &   $~~$235$~~~~$    &  F0 \\
   J173650.63+060814.4 &  264.21098 & 6.1373510 &     $-$30.9 $\pm$ 18.7&  $-$15.5 $\pm$  7.0  &  16.57 & 15.25 & 14.73 & 14.50 &  14.39 &  13.44 &  13.03 &  12.99 &  12.95 &  $-$1.51 &   $-$436$~~~~$   &  $-$416$~~~~$    &  F5 \\
   J224207.91+072808.3 &  340.53297 & 7.4689990 &   60.2 $\pm$  2.3&  $-$42.7 $\pm$  2.6  &  16.06 & 15.28 & 16.33 & 14.60 &  14.43 &  13.50 &  13.18 &  13.24 &  13.15 &  $-$2.18 &   $-$462$~~~~$   &  $-$436$~~~~$         &  A0 \\
   J091849.92-005331.5 &  139.70801 &$-$0.892091& $-$45.3 $\pm$  3.2&  $-$85.7 $\pm$  3.0  &  14.79 & 13.91 & 13.73 & 13.47 &  13.42 &  12.48 &  12.24 &  12.20 &  12.15 &  $-$1.81 &   $~~$453$~~~~$   &   $~~$465$~~~~$     &  A5 \\
   J122723.65+482136.1 &  186.84858 & 48.360029 &  15.3 $\pm$  3.5&  $-$88.2 $\pm$  3.7  &  17.04 & 15.88 & 15.44 & 15.28 &  15.19 &  14.36 &  13.96 &  13.84 &  13.89 &  $-$1.11 &   $-$209$~~~~$   &  $-$213$~~~~$          &  G0 \\
   J124330.95+423119.9 &  190.87898 & 42.522216 &  $-$25.4 $\pm$  2.5&  $-$55.5 $\pm$  2.5  &  16.74 & 15.90 & 15.64 & 15.56 &  15.53 &  14.85 &  14.56 &  14.46 &  14.55 &  $-$2.20 &   $-$229$~~~~$   &  $-$263$~~~~$       &  A0 \\
   J121437.93+542522.9 &  183.65808 & 54.423052 &   17.0 $\pm$  3.8&  $-$62.6 $\pm$  4.0  &  15.93 & 14.98 & 14.72 & 14.63 &  14.65 &  13.82 &  13.59 &  13.54 &  13.57 &---$~~$ &   $-$203$~~~~$   &  $-$197$~~~~$           &  A0 \\
   J212028.30-013323.4 &  320.11794 &$-$1.556502 & 51.2 $\pm$  3.8&  $-$47.0 $\pm$  4.0  &  16.87 & 15.99 & 15.67 & 15.54 &  15.53 &  14.76 &  14.36 &  14.54 &  14.30 &  $-$2.18 &   $-$338$~~~~$   &  $-$311$~~~~$          &  F0 \\
   J080608.76+063349.8 &  121.53652 & 6.5638360 &   14.2 $\pm$  4.6&  $-$20.4 $\pm$  5.1  &  16.84 & 15.91 & 15.65 & 15.57 &  15.54 &  14.84 &  14.50 &  14.43 &  14.64 &  $-$1.25 &   $~~$427$~~~~$   &   $~~$423$~~~~$      &  A0 \\
   J065436.50+170313.8 &  103.64375 & 17.053856 &  $-$1.4 $\pm$  2.1&  $-$16.0 $\pm$  2.3  &  15.57 & 14.05 & 13.55 & 13.40 &  13.60 &  12.31 &  11.98 &  11.91 &  11.84 &  $-$1.23 &   $~~$290$~~~~$   &   $~~$307$~~~~$     &  F9 \\
   J102757.74+090030.0 &  156.99059 &  9.008337 &  $-$160.0 $\pm$  8.0&  $-$35.0 $\pm$  8.0  &  15.67 & 14.50 & 14.03 & 14.06 &  13.73 &  12.28 &  12.29 &  12.80 &  12.38 &---$~~$ &   $~~$293$~~~~$   &    $~~$314$~~~~$    &  G2 \\
   J120758.23+093231.9 &  181.99265 &  9.542214 &  $-$100.0 $\pm$  8.0&  $-$137.0 $\pm$  8.0  &  15.83 & 14.60 & 14.00 & 14.24 &  13.65 &  12.04 &  12.07 &  12.65 &  12.10 &  $-$1.55 &   $~~$215$~~~~$  &     $~~$214$~~~~$ &  G2 \\

\hline \multicolumn{18}{l}{{\sc Notes:} $\mu_{\alpha}$ cos($\delta$)
  and $\mu_{\delta}$ were taken from the UCAC4; $u, g, r, i$, and $z$
  are from SDSS photometry; $J, H$, and $K_{\rm s}$ are taken from the
  Two Micron All-Sky Survey (2MASS) catalog; $W2$ is from the WISE catalog; $V^{a}_{\rm r}$ was
  obtained from the LAMOST}\\
\multicolumn{18}{l}{  1D pipeline,while $V^{b}_{\rm r}$ is based on our IRAF reduction; SpTy is the spectral type
  from the LAMOST 1D pipeline.}
\end{tabular}
\end{table*}


\begin{table*}
 \centering
\caption{Spatial Positions and Kinematic Parameters of the 28 HVS
  Candidates.}
\label{tab2}
\tiny
\begin{tabular}{ccccrcrrrrrrrr}
\\
\hline
  Name  & Designation$~~$&    R.A.     &   Dec.      &     $A_{V}$$~$  &    $D_\odot$  &  $r_{\rm gc}$  &  $x$$~~$  &    $y$$~~$ &    $z$$~~$  &    $V_x$$~~~~~$  &     $V_y$$~~~~~$   &    $V_z$$~~~~~$  & $V_{\rm gt}$$~~~~$ \\
 $~$ & $~$ &deg & deg & mag$~$ & kpc& kpc & kpc & kpc & kpc & $~~~~$km s$^{-1}$ & $~~~~$km s$^{-1}$& $~~~~$km s$^{-1}$&  $~~$km s$^{-1}$ \\
\hline
LMST\_HVS1  &  J085819.90+150352.4 & 134.58293 & 15.064578  & 0.062  & 3.3 &10.5   &$-$10.3  & $-$1.5 &    1.9  &  $-$227 $\pm$  7 &  $-$705 $\pm$ 42 &  $-$288 $\pm$  6  &  795 $\pm$ 35   \\
LMST\_HVS2  &  J092707.09+242752.9 & 141.77954 & 24.464709  & 0.085  & 2.3 & 9.6   & $-$9.5  & $-$0.6 &    1.6  &  $-$655 $\pm$ 10 &     92 $\pm$ 79 &  $-$233 $\pm$ 45  &  707 $\pm$  7   \\
LMST\_HVS3  &  J074233.15+322433.8 & 115.63816 & 32.409389  & 0.104  & 0.4 & 8.4   & $-$8.3  & $-$0.0 &    0.1  &  $-$329 $\pm$ 18 &  $-$412 $\pm$ 51 &  $-$382 $\pm$ 24  &  652 $\pm$ 56   \\
LMST\_HVS4  &  J225922.58+262409.9 & 344.84410 & 26.402775  & 0.256  & 0.9 & 8.1   & $-$8.0  &    0.8 & $-$0.5  &      46 $\pm$ 26 &    248 $\pm$ 34 &     470 $\pm$ 36  &  535 $\pm$ 46  \\
LMST\_HVS5  &  J115445.93+523520.3 & 178.69142 & 52.588984  & 0.051  & 1.1 & 8.4   & $-$8.4  &    0.3 &    1.0  &     492 $\pm$ 41 &   $-$50 $\pm$  2 &      48 $\pm$ 31  &  497 $\pm$ 43  \\
LMST\_HVS6  &  J082932.48+282613.3 & 127.38536 & 28.437039  & 0.085  & 0.2 & 8.2   & $-$8.2  & $-$0.0 &    0.1  &  $-$311 $\pm$ 15 &  $-$318 $\pm$ 42 &  $-$177 $\pm$ 15  &  480 $\pm$ 43   \\
LMST\_HVS7  &  J231232.05+265039.5 & 348.13358 & 26.844326  & 0.171  & 1.6 & 8.3   & $-$8.1  &    1.4 & $-$0.8  &  $-$465 $\pm$ 82 &   $-$55 $\pm$  0 &      10 $\pm$  6  &  469 $\pm$ 81  \\
LMST\_HVS8  &  J135931.65+413121.6 & 209.88189 & 41.522689  & 0.033  & 1.0 & 8.0   & $-$7.9  &    0.3 &    0.9  &      31 $\pm$  0 &  $-$164 $\pm$ 22 &     409 $\pm$ 22  &  443 $\pm$ 28   \\
LMST\_HVS9  &  J134621.50+301548.2 & 206.58962 & 30.263401  & 0.045  & 2.0 & 7.9   & $-$7.7  &    0.3 &    2.0  &     238 $\pm$ 29 &  $-$326 $\pm$  8 &  $-$129 $\pm$  8  &  425 $\pm$ 11   \\
LMST\_HVS10 &  J110208.65+575200.2 & 165.53608 & 57.866742  & 0.024  & 0.8 & 8.4   & $-$8.4  &    0.2 &    0.6  &     409 $\pm$ 28 &   $-$56 $\pm$  9 &      85 $\pm$ 37  &  423 $\pm$ 35  \\
LMST\_HVS11 &  J224145.15+292428.4 & 340.43813 & 29.407892  & 0.192  & 1.4 & 8.1   & $-$8.0  &    1.3 & $-$0.6  &     326 $\pm$ 49 &   $-$76 $\pm$ 11 &  $-$171 $\pm$  5  &  377 $\pm$ 43   \\
LMST\_HVS12 &  J124020.02+453259.6 & 190.08345 & 45.549896  & 0.042  & 1.4 & 8.4   & $-$8.2  &    0.3 &    1.3  &     373 $\pm$ 36 &   $-$10 $\pm$  7 &   $-$58 $\pm$ 19  &  379 $\pm$ 32   \\
LMST\_HVS13 &  J123134.97+424736.1 & 187.89571 & 42.793379  & 0.073  & 1.3 & 8.3   & $-$8.2  &    0.2 &    1.3  &     316 $\pm$ 30 &   $-$32 $\pm$  6 &  $-$152 $\pm$ 18  &  354 $\pm$ 18   \\
LMST\_HVS14 &  J094122.37-000822.2 & 145.34323 &$-$0.139505 & 0.167  & 0.9 & 8.4   & $-$8.4  & $-$0.6 &    0.5  &  $-$189 $\pm$  2 &  $-$291 $\pm$ 16 &   $-$10 $\pm$  4  &  348 $\pm$ 14  \\
LMST\_HVS15 &  J073050.71+293212.6 & 112.71130 & 29.536856  & 0.189  & 0.8 & 8.7   & $-$8.7  & $-$0.1 &    0.2  &     321 $\pm$  3 &   $-$24 $\pm$ 25 &  $-$117 $\pm$ 16  &  344 $\pm$  7   \\
LMST\_HVS16 &  J123641.48+443715.9 & 189.17285 & 44.621092  & 0.045  & 1.7 & 8.5   & $-$8.3  &    0.3 &    1.6  &     116 $\pm$ 13 &  $-$294 $\pm$ 11 &  $-$116 $\pm$ 18  &  337 $\pm$  7  \\
LMST\_HVS17 &  J085623.74+211322.3 & 134.09893 & 21.222872  & 0.078  & 1.0 & 8.7   & $-$8.7  & $-$0.3 &    0.6  &  $-$292 $\pm$ 15 &   $-$50 $\pm$ 13 &  $-$132 $\pm$  5  &  325 $\pm$ 17   \\
LMST\_HVS18 &  J173650.63+060814.4 & 264.21098 & 6.1373510  & 0.217  & 0.4 & 7.6   & $-$7.6  &    0.2 &    0.1  &  $-$313 $\pm$  5 &   $-$17 $\pm$ 32 &   $-$92 $\pm$ 19  &  329 $\pm$  5   \\
LMST\_HVS19 &  J224207.91+072808.3 & 340.53297 & 7.4689990  & 0.316  & 0.8 & 7.9   & $-$7.8  &    0.6 & $-$0.5  &  $-$184 $\pm$ 23 &  $-$253 $\pm$  4 &      92 $\pm$ 29  &  329 $\pm$  8   \\
LMST\_HVS20 &  J091849.92-005331.5 & 139.70801 & $-$0.892091& 0.077  & 0.5 & 8.2   & $-$8.2  & $-$0.3 &    0.2  &  $-$202 $\pm$  2 &  $-$239 $\pm$ 18 &      76 $\pm$  2  &  322 $\pm$ 14  \\
LMST\_HVS21 &  J122723.65+482136.1 & 186.84858 & 48.360029  & 0.033  & 0.9 & 8.2   & $-$8.2  &    0.2 &    0.8  &     298 $\pm$ 26 &  $-$113 $\pm$  5 &   $-$43 $\pm$ 22  &  323 $\pm$ 22   \\
LMST\_HVS22 &  J124330.95+423119.9 & 190.87898 & 42.522216  & 0.077  & 1.5 & 8.4   & $-$8.2  &    0.3 &    1.5  &     103 $\pm$  9 &  $-$260 $\pm$ 16 &  $-$142 $\pm$ 18  &  315 $\pm$  7  \\
LMST\_HVS23 &  J121437.93+542522.9 & 183.65808 & 54.423052  & 0.051  & 1.2 & 8.4   & $-$8.4  &    0.4 &    1.1  &     302 $\pm$ 29 &   $-$85 $\pm$  9 &      14 $\pm$ 22  &  316 $\pm$ 26   \\
LMST\_HVS24 &  J212028.30-013323.4 & 320.11794 & $-$1.556502& 0.085  & 1.1 & 7.4   & $-$7.4  &    0.7 & $-$0.6  &  $-$231 $\pm$ 25 &  $-$164 $\pm$  4 &  $-$125 $\pm$ 42  &  312 $\pm$ 33   \\
LMST\_HVS25 &  J080608.76+063349.8 & 121.53652 & 6.5638360  & 0.077  & 1.9 & 9.5   & $-$9.5  & $-$1.0 &    0.6  &  $-$170 $\pm$  6 &  $-$176 $\pm$  7 &     185 $\pm$ 64  &  312 $\pm$ 33  \\
LMST\_HVS26 &  J065436.50+170313.8 & 103.64375 & 17.053856  & 0.700  & 0.2 & 8.2   & $-$8.2  & $-$0.0 &    0.0  &  $-$273 $\pm$ 11 &    120 $\pm$  4 &      41 $\pm$  4  &  302 $\pm$  9   \\
LMST\_HVS27 &  J102757.74+090030.0 & 156.99059 &  9.008337  & 0.090  & 0.7 & 8.3   & $-$8.2  & $-$0.3 &    0.6  &  $-$525 $\pm$ 35 &  $-$154 $\pm$  4 &   $-$89 $\pm$  3  &  555 $\pm$ 34  \\
LMST\_HVS28 &  J120758.23+093231.9 & 181.99265 &  9.542214  & 0.050  & 0.8 & 8.0   & $-$8.0  & $-$0.2 &    0.7  &  $-$62  $\pm$ 4  &  $-$443 $\pm$ 26 &   $-$15 $\pm$  4  &  447 $\pm$ 25  \\
\hline
\multicolumn{14}{l}{{\sc Notes:} $r_{\rm gc}$ is the galactrocentric
  distance; $(x,y,z,V_x,V_y,V_z)$ are the 6D Galactic phase--space
  coordinates; $V_{\rm gt}$ is the total (space) velocity in the
  Galactic rest frame.}\\
\end{tabular}
\end{table*}

\begin{table*}
 \centering
\caption{Distances and Derived Kinematic Parameters of the 28 HVS Candidates from the Literature.}
\label{tab3}
\tiny
\begin{tabular}{ccccccccc}
\\
\hline
  Name$~~$&    R.A.     &   Dec.      &       $D_\odot$  &    $V_x$$~~~~~$  &     $V_y$$~~~~~$   &    $V_z$$~~~~~$  & $V_{\rm gt}$$~~~~$ & Ref.  \\
  $~$ &deg & deg  & kpc&  $~~~~$km s$^{-1}$ & $~~~~$km s$^{-1}$& $~~~~$km s$^{-1}$&  $~~$km s$^{-1}$ \\
\hline

LMST\_HVS1    & 134.58293 & 15.064578   &  1462       & $-$254$~~~~$     & $-$283$~~~~$      &  $-$  9$~~~~$     &  381$~~~~$    & 1  \\
LMST\_HVS2    & 141.77954 & 24.464709   &   ---       &          ---     &           ---     &          ---      &      ---      & --- \\
LMST\_HVS3    & 115.63816 & 32.409389   &   ---       &          ---     &           ---     &          ---      &      ---      & --- \\
LMST\_HVS4    & 344.84410 & 26.402775   &  1329       &     57$~~~~$     &    319$~~~~$      &     592$~~~~$     &  675$~~~~$    & 1 \\
LMST\_HVS5    & 178.69142 & 52.588984   &  1010       &    444$~~~~$     &  $-$26$~~~~$      &      21$~~~~$     &  445$~~~~$    & 1 \\
LMST\_HVS6    & 127.38536 & 28.437039   &   ---       &          ---     &           ---     &          ---      &      ---      & --- \\
LMST\_HVS7    & 348.13358 & 26.844326   &   ---       &          ---     &           ---     &          ---      &      ---      & --- \\
LMST\_HVS8    & 209.88189 & 41.522689   &  1890(714)  &40 (28)$~~~~$     & $-$571($-$19 )    &     557 (357)    &  799 (359)    & 1(2)  \\
LMST\_HVS9    & 206.58962 & 30.263401   &  2475       &    287$~~~~$     & $-$436$~~~~$      &  $-$117$~~~~$     &  536$~~~~$    & 1  \\
LMST\_HVS10   & 165.53608 & 57.866742   &   130       &    167$~~~~$     &    126$~~~~$      &  $-$135$~~~~$     &  250$~~~~$    & 1 \\
LMST\_HVS11   & 340.43813 & 29.407892   &   ---       &          ---     &           ---     &          ---      &      ---      & --- \\
LMST\_HVS12   & 190.08345 & 45.549896   &  2468       &    611$~~~~$     & $-$158$~~~~$      &   $-$30$~~~~$     &  632$~~~~$    & 1  \\
LMST\_HVS13   & 187.89571 & 42.793379   &  2378       &    493$~~~~$     & $-$185$~~~~$      &   $-$84$~~~~$     &  534$~~~~$    & 1  \\
LMST\_HVS14   & 145.34323 &$-$0.139505  &  1484(1267) & $-$195 (-193)    & $-$431 ($-$375)   &  $-$169 ($-$106)  &  503(435)     & 1(2) \\
LMST\_HVS15   & 112.71130 & 29.536856   &   798       &    321$~~~~$     &  $-$24 $~~~~$     &  $-$118$~~~~$     &  343$~~~~$    & 2  \\
LMST\_HVS16   & 189.17285 & 44.621092   &   ---       &          ---     &           ---     &          ---      &      ---      & --- \\
LMST\_HVS17   & 134.09893 & 21.222872   &  1031       & $-$293$~~~~$     & $-$52$~~~~$      &  $-$135$~~~~$     &  327$~~~~$    & 1  \\
LMST\_HVS18   & 264.21098 & 6.1373510   &   ---       &          ---     &           ---     &          ---      &      ---      & --- \\
LMST\_HVS19   & 340.53297 & 7.4689990   &  3658       & $-$565$~~~~$     &$-$830$~~~~$      &  $-$597$~~~~$     & 1169$~~~~$    & 1  \\
LMST\_HVS20   & 139.70801 & $-$0.892091 &  1455(596)  & $-$157($-$199)   & $-$523($-$263)    &  $-$244(48)       &  599(333)     & 1(2) \\
LMST\_HVS21   & 186.84858 & 48.360029   &  5099       &   1322$~~~~$     &$-$1394$~~~~$      &     608$~~~~$     & 2015$~~~~$    & 1  \\
LMST\_HVS22   & 190.87898 & 42.522216   &  2422       &    129$~~~~$     & $-$500$~~~~$      &   $-$86$~~~~$     &  524$~~~~$    & 1 \\
LMST\_HVS23   & 183.65808 & 54.423052   &  1204       &    292$~~~~$     & $-$78$~~~~$      &       7$~~~~$     &  303$~~~~$    & 1  \\
LMST\_HVS24   & 320.11794 & $-$1.556502 &   ---       &          ---     &           ---     &          ---      &      ---      & --- \\
LMST\_HVS25   & 121.53652 & 6.5638360   &  2415       & $-$138$~~~~$     & $-$220$~~~~$      &     187$~~~~$     &  320$~~~~$    & 1 \\
LMST\_HVS26   & 103.64375 & 17.053856   &   ---       &          ---     &           ---     &          ---      &      ---      & --- \\
LMST\_HVS27   & 156.99059 &  9.008337   &  1878       &$-$1143$~~~~$     & $-$494$~~~~$      &  $-$594$~~~~$     & 1380$~~~~$    & 1 \\
LMST\_HVS28   & 181.99265 &  9.542214   &  1698       & $-$147$~~~~$     &$-$1113$~~~~$      &  $-$263$~~~~$     & 1153$~~~~$    & 1 \\
\hline
\multicolumn{9}{l}{{\sc References:} 1 -- \citet{2010PASP..122.1437P}; 2 -- \citet{2011MNRAS.413.1581G}.}\\
\end{tabular}
\end{table*}

\subsection{Distance Derivation}

To obtain the six-dimensional (6D) phase--space coordinates of the HVS
candidates, we first adopted the photometric parallax relations of
\citet{2008ApJ...684..287I} to calculate the candidates' absolute
$r$-band magnitudes, $M_{r}$. For objects with $0.2< (g-i) < 4.0$ mag,
we used
\begin{equation}
M_{r}(g-i,{\rm [Fe/H]})= M_{0}+\Delta M_{r},
\label{absmag}
\end{equation}
where $\Delta M_{r}$ and $M_{0}$ are given by, respectively, Eqs (A2)
and (A7) of \citet{2008ApJ...684..287I}. Where available, we used the
[Fe/H] values derived by the LAMOST pipeline. For objects without
LAMOST metallicity determinations, photometric metallicities were
adopted based on Eq. (4) of \citet{2008ApJ...684..287I}, for $0.2 <
(g-r) < 0.6$ mag and $0.8 < (u-g) < 1.3$ mag. For stars that did not
meet either of these color criteria, but which had $(r-i)$ colors in
the range $0.1 < (r-i) < 1.6$ mag, a ``bright'' photometric parallax
relation based on Eq. (2) of \citet{2008ApJ...673..864J} was used.

To derive the appropriate extinction corrections, we used the
Rayleigh--Jeans color-excess method \citep{majewski11}, which
estimates reddening values on a star-by-star basis using a combination
of near- and mid-IR data from the Two Micron All-Sky Survey's (2MASS)
Point Source Catalog and the WISE catalog. Specifically, we used data
in the $H$ and WISE $W2$ (4.6 $\mu$m) bands:
\begin{equation}
A(K_{\rm s})= 0.918 \{ H-[4.6\;\mu {\rm m}]-(H-[4.6\;\mu {\rm m}])_{0}
\}.
 \label{A(Ks)1}
\end{equation}
We adopted $(H-4.6\;\mu {\rm m})_{0}=0.05$ mag \citep{zasowski13} and
$E(J-K_{\rm s})=1.5A_{K_{\rm s}}$ \citep{2005ApJ...619..931I}.

To avoid reddening overcorrections of halo targets, we used the
\cite[][hereinafter SFD]{schlegel98} reddening maps as upper limit to
the reddening toward stars in fields at Galactic latitudes $|b| \geq
16^\circ$ \citep[cf.][]{zasowski13}, adopting
\begin{equation}
A_{K_{\rm s}}=0.302 E(B-V)_{\rm SFD},
 \label{A(Ks)2}
\end{equation}
as long as the $E(J-K_{\rm s})$ value calculated based on the 2MASS
and WISE data was greater than 1.2$\times$ the SFD-derived value
\citep{zasowski13}. The visual extinction $A_V$ can then be obtained
from $A_{K_{\rm s}} / A_V =0.118$ \citep{1989ApJ...345..245C}. For
negative $A_{K_{\rm s}}$ values, we adopted the reddening values from
the integrated SFD maps. We derived the $r$-band extinction, $A_r$,
using $A_r/A_V=R_r/R_V=0.8$, where $R_r=2.31$ is given by
\citet{2013MNRAS.430.2188Y} and $R_V=3.1$ \citep{1999PASP..111...63F}.

The distances thus derived from the de-reddened $r$-band distance
modulus, combined with the positions of our 28 HVS candidates in 6D
phase--space, are sufficiently accurate and reliable to support our
identification of these objects as genuine HVSs. We illustrated this
in Section \ref{sect:rvver}, where we discussed the properties of the
highest-velocity star in our sample in the context of our full
database of observed and derived parameters.  In Section
  \ref{sect:discussion} we provide a number of examples of candidate
  stars whose spectra exhibit features as expected from their
  designations in the literature.

\subsection{Spatial Velocity Calculations}

We converted the RV, distance, and proper motions of each HVS
candidate to $V_x, V_y$, and $V_z$ in the (Cartesian) Galactic
coordinate system centered on the GC, where the $x$ axis points to the
direction opposite to that of the Sun, the $y$ axis is aligned with
the direction of Galactic rotation, and the $z$ axis points toward the
North Galactic Pole. The Sun is located at $x = -7.8$ kpc
\citep{mcmillan10}. The velocity of the local standard of rest (LSR)
is set at 220 km s$^{-1}$, and the motion of the Sun with respect to
the LSR is $(U_0, V_0, W_0) = (11.1, 12.24, 7.25)$ km s$^{-1}$
\citep{schonrich09}. We used the Markov Chain Monte Carlo
  technique to extract the $U, V$, and $W$ components of all
  candidates from the normally distributed parameter space of RV,
  distance, and proper motions. We estimated the typical uncertainty
  associated with the RVs based on LAMOST spectra at 13 km s$^{-1}$
  \citep{luo12}. The resulting photometric distance accuracy is then
  approximately 10\%, assuming r.m.s. values of the absolute $M_{r}$
  magnitudes of approximately 0.05 mag \citep{2008ApJ...684..287I} and
  photometric $r$-band magnitude uncertainties of $\sim 0.02$ mag
  \citep{2000AJ....120.1579Y}. The kinematic parameters pertaining to
  our 28 HVS candidates thus derived are included in Table 2.

\section{Conclusions}
\label{sect:discussion}

In this {\it Letter}, we have reported the discovery of a sample of 28
HVSs selected from the LAMOST DR1, of which 12 objects are the most
likely HVS candidates. Our sample of HVS candidates covers a much
broader color range than the equivalent ranges discussed in previous
studies and comprises the first and largest sample of HVSs in the
solar neighborhood.

We have access to sufficiently accurate observed and derived
parameters for all 28 HVS candidates to ascertain their nature as
genuine HVSs. To further verify the reliability of the sample of 28
HVSs, we matched our candidates with other catalogs. We first checked
their proper motions based on the PPMXL catalog
\citep{2010AJ....139.2440R} and did not find any significant
differences between the PPMXL and UCAC4 catalogs for these 28
HVSs. Two candidates in the New Luyten Two Tenths (NLTT) catalog
(Luyten 1979a,b), LMST\_HVS3 and LMST\_HVS6, have proper motions of
[$\mu_{\alpha} \cos(\delta), \mu_{\delta}$] = ($-154.2, -359.1$) mas
yr$^{-1}$ and ($-197.2, -384.9$) mas yr$^{-1}$, respectively
\citep{2003ApJ...582.1011S}. LMST\_HVS3 is identified as a K-type
star in the NLTT catalog, where it is catalogued as object 309-34;
careful analysis of our LAMOST spectra also classifies it as a
K-dwarf star. (The nearest NLTT star, is located at a
projected distance of approximately $12'$, so that there can be no
confusion as to our candidate HVS's nature on this basis.) We have
double checked the spectral features of all 28 HVS candidates, none
of which exhibit characteristics of white dwarfs or binary systems.
LMST\_HVS1 is also listed as a high proper-motion star [$\mu_{\alpha}
\cos(\delta) =-20.8$ mas yr$^{-1}$, $\mu_{\delta} =-54.8$ mas
yr$^{-1}$] in the catalog of stars with high proper motions
\citep{2008KFNT...24..480I}. For all three stars, the similarity of
the proper motions indicates the reliability of our kinematic data.

Since the uncertainty in distance is the major contributor to
the uncertainties in the derived total (space) velocities, we use
additional available photometric distances from
\citet{2010PASP..122.1437P} and \citet{2011MNRAS.413.1581G} to
recalculate the total velocities of our sample of 28 HVSs. Based on
these literature-based distances, the derived total velocities and
($U,V,W$) components in the Galactic rest frame are listed in
Table~\ref{tab3}, while the proper motions and RVs used are included
in Table~\ref{tab1}. The space velocities of the 28 candidates
included in Table~\ref{tab3} suggest that the candidates we
discovered are highly probable HVSs because of their intrinsically
large proper motions and RVs.

The results reported here also highlight the great potential of
discovering statistically large numbers of HVSs of different spectral
types in LAMOST survey data. We will continue to perform further
systematic HVS searches based on the LAMOST data; the resulting
stellar samples will eventually enable us to better understand the
nature of the HVSs themselves and ultimately constrain the structure
of the Galaxy.

\begin{acknowledgements}
This work was supported by ``973 Program'' 2014 CB845702, the
Strategic Priority Research Program ``The Emergence of Cosmological
Structures'' of the Chinese Academy of Sciences (CAS; grant
XDB09010100) and by the National Natural Science Foundation of China
(NSFC) under grants 11373054, 11073038, 11073001, and 11373010. The
Guoshoujing Telescope (LAMOST) is a National Major Scientific Project
built by the CAS. Funding for the project has been provided by the
National Development and Reform Commission. LAMOST is operated and
managed by the National Astronomical Observatories, CAS.
\end{acknowledgements}

\label{lastpage}
\end{document}